\documentclass[twocolumn,aps,pra]{revtex4-1}
\usepackage{times}
\usepackage{t1enc}
\usepackage[english]{babel}
\usepackage{graphicx}

\begin{document}

\title{Localization length calculations in alternating
metamaterial-birefringent disordered layered stacks}

\author{O. del Barco$^{1}$, V. Gasparian$^{2}$ and Zh. Gevorkian$^{3,4}$}
\affiliation{$^1$ Departamento de F\'{i}sica - CIOyN, Universidad de Murcia, Spain\\
$^{2}$ California State University, Bakersfield, USA\\
$^{3}$ Yerevan Physics Institute, Alikhanian Brothers St. 2, 0036 Yerevan, Armenia. \\
$^{4}$ Institute of Radiophysics and Electronics, Ashtarak-2,
0203, Armenia.\\}

\begin{abstract}
A detailed theoretical and numerical analysis of the localization
length in alternating metamaterial-birefringent random layered
stacks, under uncorrelated thickness-disorder, has been performed.
Similar structures have recently been reported to suppress the
Brewster delocalization for $\it{p}$-polarized light, when
''standard'' isotropic layers (with positive index of refraction)
are considered instead of metamaterial layers, providing a generic
means to produce polarization-insensitive, broadband reflections.
However, this enhancement of localization is valid for short
wavelengths $\lambda$ compared to the mean layer thickness $a_0$.
At higher wavelengths, we recover the Brewster anomalies for
$\it{p}$-polarized states impeding a remarkable localization of
light. To achieve a better localization for a wider range of
wavelengths, we replaced the conventional isotropic layers by
negative-index metamaterials presenting low losses and constant
index of refraction over the near-infrared range. As a result, our
numerical calculations exhibit a linear dependence of the
localization length with $\lambda$ (in the region $5 <\lambda/a_0
< 60$) reducing the Brewster anomalies in more than two orders of
magnitude with respect to the standard isotropic scheme at oblique
incidence. This enhancement of localization is practically
independent of the thickness disorder kind and is also held under
weak refractive-index disorder.
\end{abstract}


\maketitle

\section{Introduction}

Broadband reflections from a disordered dielectric medium are a
physical manifestation of the localization of light \cite{BE97}.
It has been reported the design of such high-performance broadband
mirrors with a broader reflection band than periodic systems with
the same refractive indices \cite{ZH95,LI00}. If a stack system
could be devised to effectively localize both $\it{s}$- and
$\it{p}$-polarized light over all angles of incidence, it would
provide a means to obtain polarization-insensitive, broadband
reflections with potential applications that include waveguides
and thermoelectric devices, among others \cite{FI98}. However, a
fundamental problem arises due to the so-called Brewster
delocalization \cite{AR90,DU97,LE11}, that is, the impossibility
of achieving a total localization of light at certain oblique
incidences for $\it{p}$-polarization. The main reason is that the
depth of penetration for $\it{p}$ states increases several orders
of magnitude as compared with the depth of penetration of $\it{s}$
states \cite{SI88}. This is an important issue to take into
account for potential polarization-insensitive broadband mirrors.

Recently, Jordan \emph{et al.} \cite{JO13} reported a suppression
of Brewster delocalization in stacks of alternating
isotropic-birefringent random layers under uncorrelated thickness
disorder. Stacks containing a mixture of positive uniaxial and
negative uniaxial birefringent layers have been claimed as
effective media to inhibit the Brewster delocalization for over
all angles of incidence. This work was motivated by a previous
analysis of the nonpolarizing reflections from birefringent
guanine and isotropic cytoplasm multilayer structure found in some
species of silvery fish \cite{JO12}. These authors mimic the
guanine-cytoplasm structure in fish skin where there are two
different types of birefringent crystal present with isotropic
cytoplasm gaps and obtain a noticeable enhancement of
localization. Nevertheless, as we will show briefly, this
enhancement of localization for $\it{p}$-polarized light is valid
for short wavelengths compared to the mean layer thickness. At
higher wavelengths, we recover the Brewster anomalies for $\it{p}$
states at oblique incidence, impeding a remarkable localization of
light.

In order to achieve an improvement of localization, we propose a
stack composed of alternating metamaterial-birefringent disordered
layers where the isotropic layers possess a negative index of
refraction (that is, we are dealing with left-handed (LH)
isotropic media \cite{VE68,SM00}). These artificial materials have
been reported to resolve images beyond the diffraction limit
\cite{PEN00,SMITH04}, act as an electromagnetic cloak
\cite{LEO06,SCHU06}, enhance the quantum interference
\cite{YANG08} or yield to slow light propagation \cite{PAPA09}.
Moreover, recent research on LH metamaterials formed by hybrid
metal-semiconductor nanowires \cite{PAN13} has described negative
indices of refraction in the near-infrared with values of the real
part well below -1 and extremely low losses (an order of magnitude
better than present optical negative-index metamaterials). With
this choice, we demonstrate a reduction of Brewster anomalies in
more than two orders of magnitude with respect to the standard
isotropic scheme at higher wavelengths, providing a means to
obtain polarization-insensitive, broadband reflections in a wider
wavelength region. Furthermore, this enhancement of localization
is practically independent of the thickness disorder kind and is
also held under weak refractive-index disorder.

The plan of the work is as follows. In Sec.\ II we carry out an
exhaustive description of our system and the theoretical framework
to derive an analytical expression for the localization length
$\xi_{s,p}$ in the short-wavelength regime (that is, when the
incident wavelength $\lambda$ is less than the mean layer
thickness $a_0$). A detailed numerical analysis concerning the
localization length for two different disorder distributions (the
Poisson-thickness disorder and the uniform-thickness distribution)
will be performed in Sec.\ III. Two wavelength regions will be
analyzed in detail: the previously mentioned short-wavelength
regime and an intermediate-wavelength region ($5<\lambda/a_0<60$)
where, as we will show, the improvement of localization is more
pronounced. For completeness, weak refractive-index disorder (with
random variations in the indices of refraction of each layer) is
also considered in our model. A set of analytical expressions for
$\xi_{s,p}$ in the so-called intermediate-wavelength region, as a
function of the wavelength $\lambda$ and the incident angle
$\theta$, are also derived. Finally, we summarize our results in
Sec.\ IV.

\section{System description and theoretical model}

\begin{figure}
\includegraphics[width=.45\textwidth]{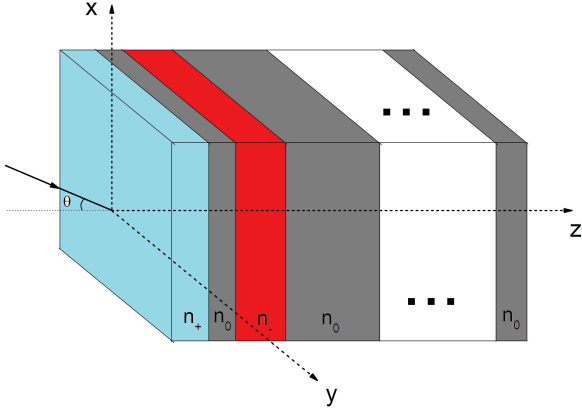}
\caption{Our disordered layered stack of
alternating uniaxial positive ($\mathbf{n}_{+}$) and uniaxial
negative ($\mathbf{n}_{-}$) layers. Gray regions correspond to
left-handed isotropic layers with negative index of refraction
$n_0$. The principal axes of the birefringent layers are aligned
to $x$ and $y$ directions, while light enters from the left at
angle $\theta$ and propagates through $z$ axis.}\label{fig1}
\end{figure}

Our structure of interest is formed by a mixture of alternating
uniaxial birefringent layers and LH isotropic media (see Fig.\
\ref{fig1}). For positive uniaxial layers, the corresponding
refractive index vector is given by $\mathbf{n}_{+} =
(n_1,n_1,n_2)$ while for negative uniaxial $\mathbf{n}_{-} =
(n_2,n_2,n_1)$. The principal axes of birefringent layers are
aligned to $x$ and $y$ directions and the ratio of negative
uniaxial to the total number of layers has been defined via the
parameter $f$. Each birefringent layer is sandwiched between two
left-handed isotropic layers with negative index of refraction
$n_0$ and the whole systems is embedded in a ''standard''
isotropic media with positive refractive index $|n_0|$. The
following condition $n_2>n_1>|n_0|$ is hold to avoid critical
angles, that is, to be sure that light is able to access all
angles of incidence at each isotropic-birefringent interface in
the stack. Light enters from the left at angle $\theta$ and
propagates through $z$ axis.

In order to effectively improve the localization of light in such
mixed structures, we need LH metamaterials with extremely low
losses and constant values of the negative-index over a relatively
wide spectrum. Such metamaterials have been recently reported by
Paniagua-Dominguez \emph{et al.} \cite{PAN13} and might operate in
the near-infrared region (from 700 to 1900 nm in this case). As we
will show in the next section, the enhancement of localization in
our system is mainly achieved in a region where the incident
wavelength $\lambda$ satisfies the inequality $20<\lambda/a_0<60$,
where $a_0$ corresponds to the mean layer thickness. Accordingly,
our mixed disordered stack can adequately be tailored to operate
in the near-infrared provided that $a_0 \simeq 35 nm$.

Let us now perform a theoretical study of the localization length
for both $\it{s}$ and $\it{p}$ states in such structures under
uncorrelated thickness disorder. To this aim, we derive an
expression for the transmission coefficient of the whole
arrangement $T_{s,p}$ and then we use the standard definition of
the localization length
\begin{equation}\label{llspo}
\xi_{s,p} = - \frac{2L}{\langle \ln T_{s,p} \rangle},
\end{equation}
where $L=N a_0$ is the system length and $N$ the total number of
layers. The angular brackets $\langle ... \rangle$ stand for
averaging over the disorder. The theoretical method here employed
is a non-perturbative approach based on the exact calculation of
the Green's function (GF) of an electromagnetic wave in a given
electric permittivity profile \cite{AR91}. Once the GF of the
layered structure is known, the transmission coefficient $T_{s,p}$
can be derived using the Fisher-Lee relation \cite{FI81}. This GF
approach is compatible with the transfer matrix method and has
been widely used to calculate the average density of states over a
sample, the energy spectrum of elementary excitations \cite{CA97}
or the characteristic barrier tunneling time \cite{GCB96}, among
others.

In our Green's function approach, the inverse of the transmission
coefficient through our multilayer structure, $1/T_{s,p}$,
coincides with the diagonal elements of the resulting transfer
matrix and can be derived via the following relation \cite{AR91}
\begin{equation}\label{TGF}
T_{s,p} = |D_N|^{-2},
\end{equation}
where $D_N$ corresponds to the so-called characteristic
determinant
\begin{equation}\label{e:Det}
D_N=D_N^{(0)} \exp\left[-i\varphi_N\right]
\prod_{l=1}^N(1-r^2_{l-1,l})^{-1/2}.
\end{equation}
Here $\varphi_N$ represents the total phase thickness accumulated
by the incident wave along the whole sample, and is given by
\begin{equation}\label{phasetot}
\varphi_N=\sum^{N}_{l=1} \varphi_l,
\end{equation}
where $\varphi_l$ is the phase thickness of each individual layer.
For our birefringent media, $\varphi_l$ satisfy \cite{JO13}
\begin{equation}\label{0}
\varphi_l=\left\{ \left(\frac{2\pi}{\lambda}\right) a_l
n_{l,x}\sqrt{1-\left(\frac{n_0^2 \sin^2\theta}{n_{l,z}^2}\right)},
\; \mbox{p-polarization} \atop \left(\frac{2\pi}{\lambda}\right)
a_l n_{l,y}\sqrt{1-\left(\frac{n_0^2
\sin^2\theta}{n_{l,y}^2}\right)}, \; \text{s-polarization}.
\right.
\end{equation}
with $a_l$ the layer thickness and $n_{l,j} \ (j=x,y,z)$ the
corresponding indices of refraction of the positive or negative
uniaxial layers
\begin{equation}\label{indicesxy}
n_{l,x}=n_{l,y} = \left\{ n_1, \; \mbox{positive layers} \atop
n_2, \; \text{negative layers}, \right.
\end{equation}
and
\begin{equation}\label{indicesz}
n_{l,z}= \left\{ n_2, \; \mbox{positive layers} \atop n_1, \;
\text{negative layers}. \right.
\end{equation}
In our LH isotropic media, the parameters $\varphi_l$ are
calculated via the following expression
\begin{equation}\label{philiso}
\varphi_l = - \left(\frac{2\pi}{\lambda}\right) a_l |n_0|
\cos\theta,
\end{equation}
for both $\it{s}$- and $\it{p}$-polarized states. Notice that the
phase thicknesses have negative sign in this case, due to the
negativeness of the indices of refraction.

$D_N^{(0)}$ corresponds to the determinant of a tridiagonal matrix
and satisfies the following recurrence relation
\begin{equation}\label{rec}
D_l^{(0)}=A_l D_{l-1}^{(0)} - B_l D_{l-2}^{(0)},
\end{equation}
with the initial conditions
\[
A_1=1, \; D_{0}^{(0)}=1, \;  \text{and} \; D_{-1}^{(0)}=0.
\]
For $l>1$ we have
\begin{equation}
A_l=1+\frac{r_{l-1,l}}{r_{l-2,l-1}}
\exp\left[2i\varphi_{l}\right],
\end{equation}
and
\begin{equation}
B_l=(A_l-1){(1-r_{l-1,l-2}^2)},
\end{equation}
where the quantity $r_{l-1,l}$ symbolizes the Fresnel reflection
amplitude of light propagating from region $l-1$ into region $l$.
In our case, these coefficients are given by the generalized
isotropic-birefringent Fresnel relations \cite{JO13,AZ88}
\begin{equation}\label{rs2}
r_{l-1,l} = \frac{|n_0| \cos\theta -
\sqrt{n_{l,y}^2-n_0^2\sin^2\theta}}{|n_0| \cos\theta +
\sqrt{n_{l,y}^2-n_0^2\sin^2\theta}},
\end{equation}
for $\it{s}$ polarization and
\begin{equation}\label{rp2}
r_{l-1,l} = \frac{|n_0| n_{l,x} n_{l,z} \cos\theta - n_0^2
\sqrt{n_{l,z}^2-n_0^2\sin^2\theta}}{|n_0| n_{l,x} n_{l,z}
\cos\theta + n_0^2 \sqrt{n_{l,z}^2-n_0^2\sin^2\theta}},
\end{equation}
for $\it{p}$-polarized states. As it is well-known, these
reflection coefficients for $\it{p}$ polarization vanish at the
generalized anisotropic Brewster angle $\theta_{\rm B}$
\cite{OR14}
\begin{equation}\label{GBA}
\tan(\theta_{\rm B}) = \left(\frac{n_{l,z}}{|n_0|}\right)
\sqrt{\frac{n_0^2-n_{l,x}^2}{n_0^2-n_{l,z}^2}}.
\end{equation}

If the Fresnel coefficients $r^2_{l,l-1}$ are small, one can
easily find (after simple inspection of the recurrence relation
Eq.\ (\ref{rec})) that the parameter $D_N^{(0)}=1$. Therefore, the
determinant $D_N$ (and, consequently, the transmission coefficient
$T_{s,p}$) may be calculated at arbitrary distribution of the
layers thicknesses. Provided that our system presents thickness
disorder only (so the reflection coefficients $r^2_{l-1,l}$ remain
invariable along the averaging process) and for a large number of
layers $N$, we can write
\begin{equation}\label{prodrrr}
\prod_{l=1}^N(1-r^2_{l-1,l})^{-1/2} \simeq
\left[(1-r_{s_{+},p_{+}}^{2})^{N_{+}}
(1-r_{s_{-},p_{-}}^{2})^{N_{-}} \right]^{-1/2},
\end{equation}
where $N_{\pm}$ stands for the total number of positive or
negative uniaxial layers, and the reflection coefficients
$r_{s_{\pm},p_{\pm}}$ are given by Eqs.\ (\ref{rs2}) and
(\ref{rp2}), with the appropriate choice of uniaxial layer (that
is, positive or negative) and incident polarization. After some
trivial algebra, we obtain the following analytical expression for
the localization length $\xi_{s,p}$ (see again Eq.\ (\ref{llspo}))
\begin{equation}\label{llsp}
\frac{2a_0}{\xi_{s,p}} = - \ln \left[\left(1-r_{s_{+},p_{+}}^{2}
\right)^{1-f} \left(1-r_{s_{-},p_{-}}^{2} \right)^{f} \right],
\end{equation}
where $f=N_{-}/N$. Notice that, under the low-reflection
assumption, our Eq.\ (\ref{llsp}) is independent of the thickness
disorder kind and the incident wavelength. We will analyze later
the validity of this theoretical result for short wavelengths
compared to the mean layer thickness, that is, $\lambda<a_0$.

\begin{figure}
\includegraphics[width=.49\textwidth]{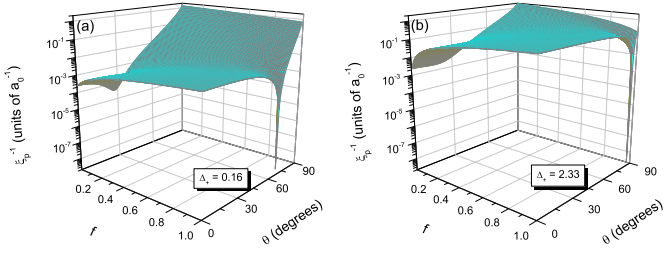}
\caption{Inverse of the localization length
$\xi_p^{-1}$ (calculated via Eq.\ (\ref{llsp})) versus the
parameter $f$ and the angle of incidence $\theta$. Two different
values of the positive layers birefringence $\Delta_{+} = n_2-n_1$
have been chosen.}\label{fig2}
\end{figure}

After simple inspection of Eq.\ (\ref{llsp}), one observes that
the Brewster anomalies for $\it{p}$-polarized states only occur
when $f=0$ (isotropic-positive stacks) or $f=1$
(isotropic-negative stacks). In such particular cases, the
right-hand side of Eq.\ (\ref{llsp}) becomes zero at $\theta_{\rm
B}$ resulting in the logarithmic divergence of the localization
length $\xi_p$ and, consequently, in the delocalization of
$\it{p}$ states. This result can be easily evidenced in Fig.\
\ref{fig2} where we represent the inverse of the localization
length for $\it{p}$-polarization, $\xi_p^{-1}$ (calculated via
Eq.(\ref{llsp})) versus the parameter $f$ and the angle of
incidence $\theta$. Two different values of the positive
birefringence $\Delta_{+} = n_2-n_1$ have been selected (with our
choice of the refractive index vectors, the following relation
holds $\Delta_{-} = -\Delta_{+}$). The index of refraction of the
left-handed layers has been chosen to be $n_0=-1.33$. One observes
that for real mixtures of positive and negative uniaxial layers
\cite{JO13} (that is, when $0<f<1$) no Brewster anomalies occur.
For higher birefringence parameters (see Fig.\ \ref{fig2}(b)) we
obtain a smoothly varying landscape for $\xi_p^{-1}$.
Consequently, as we increase the strength of the layers
birefringence, the more localized states we achieve.

\section{Numerical results}

In this section we present our numerical results concerning the
localization length for two different disorder distributions, that
is, the Poisson-thickness disorder and the uniform-thickness
distribution (the latter also under weak refractive-index
disorder). Two wavelength regions will be analyzed in detail: the
short-wavelength regime ($\lambda/a_0 \leq 1$), where we will
check the validity of our main theoretical result Eq.\
(\ref{llsp}) and an intermediate-wavelength region
($5<\lambda/a_0<60$) with an enhancement of localization up to two
orders of magnitude with respect to standard isotropic layers.

\subsection{Poisson-thickness disorder}

\begin{figure}
\includegraphics[width=.50\textwidth]{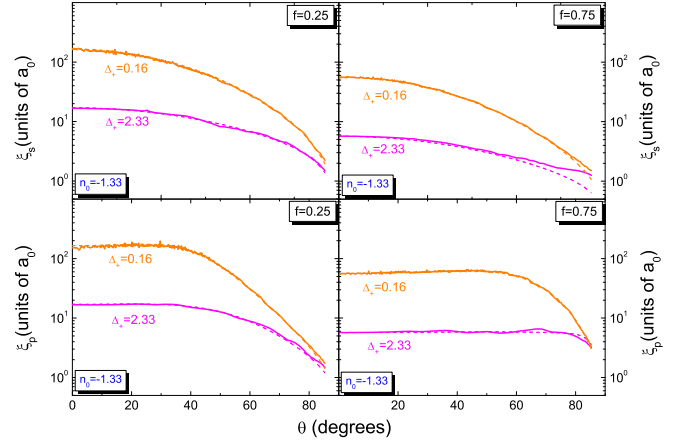}
\caption{Localization length $\xi_{s,p}$ versus the
angle of incidence $\theta$ in the short-wavelength regime
($\lambda/a_0 = 0.032$) for a Poisson-thickness distribution. Top
figures stand for $\it{s}$-polarized states while bottom figures
represent $\it{p}$-polarized light. Solid lines symbolize our
numerical calculations whereas dashed lines show our theoretical
results (Eq.\ (\ref{llsp})).}\label{fig3}
\end{figure}

Let us first consider the well-known Poisson thickness
distribution for our disordered system, where the averaging
process of $\ln T_{s,p}$ is carried out via the following
expression
\begin{equation}\label{Poiaver}
\langle \ln T_{s,p} \rangle = \frac{1}{a_0} \int_{0}^{L} da \
\exp\left[-\frac{a}{a_0}\right] \ln T_{s,p},
\end{equation}
where, for each single realization, the numerical computation of
the quantity $T_{s,p}$ is obtained via the exact recurrence
relation Eq.\ (\ref{rec}). To this aim, we show in Fig.\
\ref{fig3} the localization length $\xi_{s,p}$ versus the angle of
incidence $\theta$ in disordered stacks with mixing ratios of
$f=0.25$ (left column) and $f=0.75$ (right column) in the
short-wavelength regime ($\lambda/a_0 = 0.032$). Dashed lines
correspond to our theoretical results based on Eq.\ (\ref{llsp}).
One observes a good agreement between our numerical and
theoretical results with an error of less than 5\%. As can be
noticed from Fig.\ \ref{fig3}, in the short-wavelength regime the
localization length of $\it{p}$ states remains constant in a wide
range of incident angles (about $45$ degrees for $f=0.25$ and $70$
degrees for $f=0.75$). As previously commented, higher values of
the positive birefringence $\Delta_{+}$ yields to an enhancement
of localization. This is a manifestation of a total suppression of
Brewster delocalization, in good agreement with recent numerical
results \cite{JO13}.

\begin{figure}
\includegraphics[width=.50\textwidth]{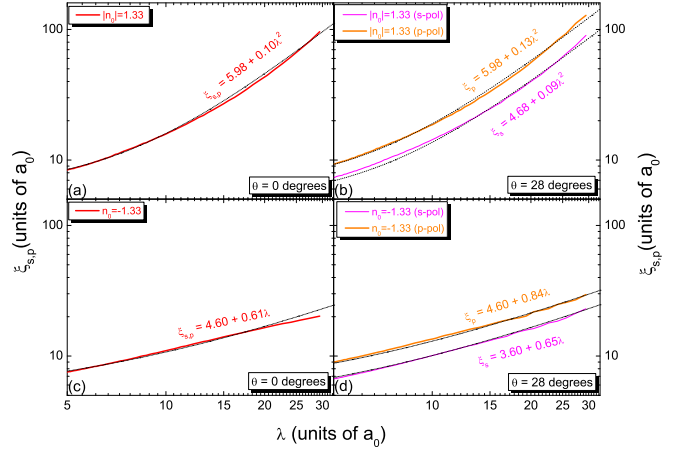}
\caption{Localization length $\xi_{s,p}$ versus the
incident wavelength $\lambda$ for two different angles of
incidence, $\theta = 0$ and $\theta = 28$ degrees. Top figures
represent ''standard'' isotropic-birefringent stacks (with
positive indices of refraction) while bottom figures stand for LH
isotropic-birefringent disordered structures. The birefringence of
the positive layers has been chosen to be $\Delta_{+}=2.33$ and
the ratio $f=0.75$ in all cases. Solid lines represent our
numerical calculations (Poisson thickness distribution, Eq.\
(\ref{Poiaver})) whereas the dotted lines symbolize the
localization length calculations performed via Eqs.\
(\ref{snormal}) and (\ref{pnormal}) (top figures) and Eqs.\
(\ref{smeta}) and (\ref{pmeta}) (bottom figures).}\label{fig4}
\end{figure}

For higher wavelengths the situation is markedly different, as can
be easily observed in Fig.\ \ref{fig4}. Here we represent our
numerical calculations for the localization length $\xi_{s,p}$ as
a function of the wavelength $\lambda$ for two different angles of
incidence, $\theta = 0$ and $\theta = 28$ degrees. Let us first
concentrate on the top figures \ref{fig4}(a) and \ref{fig4}(b)
where isotropic layers with index of refraction $|n_0|=1.33$ have
been considered. The birefringence of positive layers has been
chosen to be $\Delta_{+}=2.33$ and the ratio $f=0.75$. Solid lines
symbolize our numerical results performed via Eq.\
(\ref{Poiaver}). One notices that $\xi_{s,p}$ has a quadratic
dependence on $\lambda$ at higher wavelengths, a typical result
for various disordered models and the subject of intensive ongoing
research \cite{BA12,TO13}. Dotted lines correspond to the
following numerically-deduced equations for the localization
length as a function of the wavelength
\begin{equation}\label{snormal}
\frac{\xi_s}{a_0} = \frac{1.30 \xi_{s_{0}}}{a_0} +
\frac{1}{1.15 C^2 r_s^2 (\cos\theta)^{0.5}}
\left(\frac{\lambda}{a_0}\right)^2
\end{equation}
\begin{equation}\label{pnormal}
\frac{\xi_p}{a_0} = \frac{1.30 \xi_{p_{0}}}{a_0} +
\frac{1}{1.15 C^2 r_p^2 (\cos\theta)^{2.0}}
\left(\frac{\lambda}{a_0}\right)^2,
\end{equation}
where $C=1.15 (n_1+n_2+|n_0|)$ and $\xi_{s_{0}}$ ($\xi_{p_{0}}$)
represents the localization length of $\it{s}$ ($\it{p}$) states
in the limit of short wavelengths (see again Eq. (\ref{llsp})). In
addition, $r_{s,p}^2$ stands for the average interfacial
reflection coefficient in our system \cite{AR91,JO13} and is given
by
\begin{equation}\label{r2av}
r_{s,p}^2 = (1-f) r^2_{s_{+},p_{+}} + f r^2_{s_{-},p_{-}}.
\end{equation}
An in-deep numerical analysis has been performed to derive Eqs.\
(\ref{snormal}) and (\ref{pnormal}). To this aim, $20$ different
layered stacks at different angles of incidence have been
considered, and a detailed study of the coefficient characterizing
the quadratic dependence on $\lambda$ has been carried out. Notice
that the results derived via Eqs.\ (\ref{snormal}) and
(\ref{pnormal}) are in good agreement with our numerical
calculations for angles of incidence up to $28$ degrees.
At higher angles of incidence (not shown in our figures)
the deviation between numerical and analytical results
is greater than 5\%.

Let us now replace the ''standard'' isotropic layers by
left-handed isotropic metamaterials (this change, as previously
mentioned, assumes that the index of refraction $n_0$ is now
negative). As it was recently shown by our group for isotropic
layered structures \cite{BA12}, this change of sign results in a
significant modification of the transmission coefficient
$T_{s,p}$. As a consequence, the localization length $\xi_{s,p}$
presents a region of linear dependence with $\lambda$ when the
positions of the layer boundaries are randomly shifted with
respect to ordered periodic values. The same behavior is found for
LH isotropic-birefringent layered stacks.

The physical explanation for this effect is related to the
vanishing of the total phase accumulated along the structure. This
vanishing only occurs for stacks of alternating right-handed media
(isotropic or birefringent) and left-handed isotropic layers,
where the alternating phases possess opposite sign. Provided
strong uncorrelated thickness disorder, the total phase
accumulated over the whole structure is practically null. Thus,
the transmission coefficient $T_{s,p}$ (which strongly depends on
the total phase along the structure) has a smoother dependence on
the frequency $\omega$ than the typical right-handed stacks, where
the average of $\ln T_{s,p}$ (see the denominator of Eq.\
(\ref{llspo})) gets the well-known dependence $\omega^2$. In our
case, this average has a linear dependence with the frequency
(that is, the localization length is proportional to the
wavelength $\lambda$).

Our numerical calculations confirm this feature as can be noticed
in Figs.\ \ref{fig4}(c) and \ref{fig4}(d). Solid lines symbolize
our numerical calculations (Poisson thickness distribution, Eq.\
(\ref{Poiaver})) while dotted lines correspond to our
numerically-deduced set of equations
\begin{equation}\label{smeta}
\frac{\xi_s}{a_0} = \frac{\xi_{s_{0}}}{a_0} + \frac{2}{C |r_s|
(\cos\theta)^{0.5}} \left(\frac{\lambda}{a_0}\right)
\end{equation}
\begin{equation}\label{pmeta}
\frac{\xi_p}{a_0} = \frac{\xi_{p_{0}}}{a_0} + \frac{2}{C |r_p|
(\cos\theta)^{3.0}} \left(\frac{\lambda}{a_0}\right).
\end{equation}
After direct comparison of Figs.\ \ref{fig4}(c) and \ref{fig4}(d)
(left-handed isotropic-birefringent stacks) with Figs.\
\ref{fig4}(a) and \ref{fig4}(b) (''standard''
isotropic-birefringent stacks) one observes an enhancement of the
localization for isotropic metamaterial structures at higher
wavelengths, mainly due to a smoother dependence of the
localization length $\xi_{s,p}$ with the incident wavelength
$\lambda$.

\subsection{Uniform-thickness disorder}

Let us now study another type of thickness disorder where the
arrangement of layers has random boundaries \cite{BA12}. For
completeness, weak refractive-index disorder has also been
included in our model, where we have randomly changed the indices
of refraction of each layer via the following relations $n_0 =
n_0^{(0)} + \sigma_j \delta$ (isotropic layers) and
$\mathbf{n}_{\pm} = \mathbf{n}_{\pm}^{(0)} + \sigma_j \delta$
(birefringent layers, either positive or negative). The
superscripts $(0)$ indicate unperturbed refractive indices whereas
the parameters $\sigma_j$ are zero-mean independent random numbers
within the interval $[-0.5,0.5]$. The strength of the disorder is
measured by the parameter $\delta$ which has been chosen to be
roughly a 2\% of the corresponding unperturbed indices of
refraction (that is, we assume weak refractive-index disorder).

\begin{figure}
\includegraphics[width=.50\textwidth]{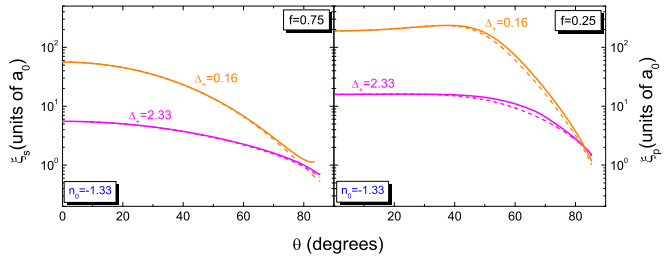}
\caption{Localization length $\xi_{s,p}$ versus the
angle of incidence $\theta$ in the short-wavelength regime
($\lambda/a_0 = 0.032$) for an uniform-thickness distribution.
Left figure stands for $\it{s}$ polarization while the right one
represents $\it{p}$-polarized light. Solid lines symbolize our
numerical calculations whereas dashed lines show our theoretical
results (Eq.\ (\ref{llsp})).}\label{fig5}
\end{figure}

Firstly, the short-wavelength regime will be considered. In Fig.\
\ref{fig5} we show the localization length $\xi_{s,p}$ versus the
angle of incidence $\theta$ in such disordered stacks with mixing
ratios of $f=0.75$ (left column) and $f=0.25$ (right column) for
$\lambda/a_0 = 0.032$. Dashed lines correspond to our theoretical
result, Eq.\ (\ref{llsp}). Even though this equation was derived
for thickness-disorder only, a good agreement between our
numerical and theoretical results is found, provided weak
refractive-index disorder.

\begin{figure}
\includegraphics[width=.48\textwidth]{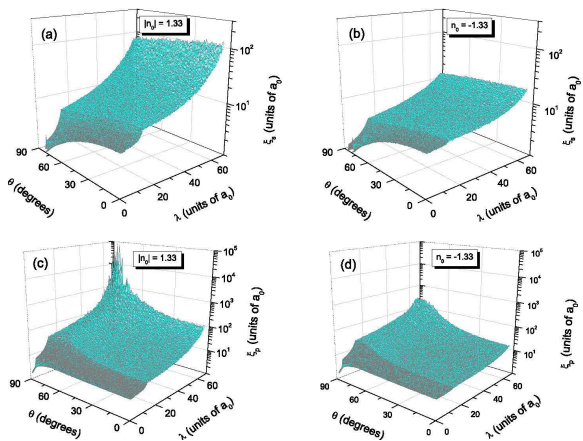}
\caption{Enhancement of localization, under a mixed
scheme of uniform-thickness and weak refractive-index disorder,
for LH isotropic-birefringent stacks. Numerical calculations for
$\xi_{s,p}$ versus the wavelength $\lambda$ and the angle of
incidence $\theta$ are shown. The birefringence of positive layers
was chosen to be $\Delta_{+}=2.33$ while the ratio $f=0.75$ in all
cases. Left-column figures represent our numerical results for
isotropic-birefringent stacks with $|n_0|=1.33$ whereas
right-column graphs stand for our left-handed isotropic stacks
with $n_0=-1.33$. The simulations were averaged over 800 random
configurations of the layered stacks.}\label{fig6}
\end{figure}

As we increase the incident wavelength, the behavior of the
localization length is significantly different (as previously
stated in the Poisson-thickness distribution). In Fig.\ \ref{fig6}
we represent three-dimensional plots of the localization length
$\xi_{s,p}$ for a combined scheme of uniform-thickness and
refractive-index disorder. Left-hand figures display the numerical
calculations for isotropic-birefringent stacks with $|n_0|=1.33$,
whereas right-column graphs stand for our left-handed isotropic
stacks with $n_0=-1.33$. In order to perform a reliable average of
the logarithm of the transmission coefficient, $\ln T_{s,p}$,
$800$ random configurations of our disordered stack have been
considered in all cases. Once more, the birefringence of positive
layers was $\Delta_{+}=2.33$ and the ratio $f=0.75$. It can be
noticed that, at long wavelengths, a decrease of one order of
magnitude for $\xi_p$ at normal incidence is achieved (compare
Fig.\ \ref{fig6}(c) to \ref{fig6}(d)) and even up to two orders of
magnitude at higher angles. Similar numerical results have been
obtained taking into account uniform-thickness disorder alone (not
shown in our figures). Consequently, under a strong uncorrelated
positional disorder scheme, weak refractive-index disorder does
not affect significatively the localization of light.

\section{Discussion and Conclusions}

A detailed theoretical and numerical analysis of the localization
length, under uncorrelated thickness-disorder, in alternating
metamaterial-birefringent random layered stacks has been
performed. Our work was motivated by the recently reported
suppression of Brewster delocalization in alternating
isotropic-birefringent stacks \cite{JO12}, where Jordan \emph{et
al.} mimic the guanine-cytoplasm structure in fish skin formed by
two different types of birefringent crystal present with isotropic
cytoplasm gaps \cite{JO13}. Nevertheless, this enhancement of
localization for $\it{p}$-polarized states is achieved for short
wavelengths (compared to the mean layer thickness $a_0$). At
higher wavelengths, we recover the Brewster anomalies for $\it{p}$
states at oblique incidence, impeding a remarkable localization of
light.

Our disordered metamaterial-birefringent stacks can effectively
improve the localization of light in the so-called
intermediate-wavelength region (in our case, $5<\lambda/a_0<60$)
up to two orders of magnitude with respect to the standard
isotropic scheme at oblique incidence. This is mainly due to the
smoother dependence of the localization length with the incident
wavelength (linear dependence) in this particular region, unlike
the well-known quadratic dependence in positive-index disordered
stacks. Moreover, this enhancement of localization is practically
independent of the thickness disorder kind and is also held under
weak refractive-index disorder.

In order to reduce losses in our system (principally due to the
presence of left-handed layers) we considered metamaterials with
extremely low losses and constant values of the negative-index
over the near-infrared region (from 700 to 1900 nm in this case)
\cite{PAN13}. As shown in Sec.\ III, the predominant enhancement
of localization in our system is achieved in the region
$20<\lambda/a_0<60$. Accordingly, our mixed disordered stack can
effectively be tailored to operate in the near-infrared provided
that $a_0 \simeq 35 nm$.

Our group consider that these metamaterial-birefringent disordered
structures might provide great insight for the experimental design
of polarization-insensitive, broadband reflectors.

\acknowledgements

The authors would like to acknowledge Miguel Ortu\~{n}o for
helpful discussions. V. G. acknowledges partial support by FEDER
and the Spanish DGI under Project No. FIS2010-16430.

\bigbreak

\end{document}